\documentclass[aps,onecolumn,superscriptaddress]{revtex4}
\usepackage{graphicx}
\usepackage{textcomp}

\begin{document}
\author{F.B.P. Niesler}
\email{fabian.niesler@kit.edu}
\affiliation{Institut f\"ur Angewandte Physik and DFG-Center for Functional Nanostructures (CFN), Karlsruhe Institute of Technology (KIT), Wolfgang-Gaede-Strasse 1, 76131 Karlsruhe, Germany}
\affiliation{Institut f\"ur Nanotechnologie, Karlsruhe Institute of Technology (KIT), 76021 Karlsruhe, Germany}
\author{J.K. Gansel}
\affiliation{Institut f\"ur Angewandte Physik and DFG-Center for Functional Nanostructures (CFN), Karlsruhe Institute of Technology (KIT), Wolfgang-Gaede-Strasse 1, 76131 Karlsruhe, Germany}
\author{S. Fischbach}
\affiliation{Institut f\"ur Angewandte Physik and DFG-Center for Functional Nanostructures (CFN), Karlsruhe Institute of Technology (KIT), Wolfgang-Gaede-Strasse 1, 76131 Karlsruhe, Germany}
\author{M. Wegener}
\affiliation{Institut f\"ur Angewandte Physik and DFG-Center for Functional Nanostructures (CFN), Karlsruhe Institute of Technology (KIT), Wolfgang-Gaede-Strasse 1, 76131 Karlsruhe, Germany}
\affiliation{Institut f\"ur Nanotechnologie, Karlsruhe Institute of Technology (KIT), 76021 Karlsruhe, Germany}
\date{\today}
\title{Metamaterial metal-based bolometers}

\begin{abstract}
We demonstrate metamaterial metal-based bolometers, which take advantage of resonant absorption in that a spectral and/or polarization filter can be built into the bolometer. Our proof-of-principle gold-nanostructure-based devices operate around 1.5 \textmu m wavelength and exhibit room-temperature time constants of about 134 \textmu s. The ultimate detectivity is limited by Johnson noise, enabling room-temperature detection of 1 nW light levels within 1 Hz bandwidth. Graded bolometer arrays might allow for integrated spectrometers with several octaves bandwidth without the need for gratings or prisms and for integrated polarization analysis without external polarization optics. 

\end{abstract}
\maketitle


Metamaterial resonances have given rise to quite a few novel material properties such as, e.g., magnetism in the optical range and negative phase velocities of light  \cite{Shalaev:2007ly,Soukoulis:2010qf,Soukoulis:2011uq}, giant circular dichroism \cite{Plum:2009bh,Zhang:2009dq,Gansel:2009cr}, or enhanced nonlinearities \cite{Klein:2006nx,Kujala:2007oq}.
Causality unavoidably connects these resonances with large imaginary parts or losses \cite{Stockman:2007zr,Kinsler:2008ve}. The associated absorption of light has often been a nuisance, especially in metal-based structures \cite{Shalaev:2007ly,Soukoulis:2010qf,Soukoulis:2011uq,Boltasseva:2011mi}. "If you cannot get rid of it, try to do something useful with it." This spirit has lately motivated the metamaterial community to search for positive sides of this absorption. For example, perfect absorbers that neither transmit nor reflect light in a certain spectral range have emerged from this search \cite{Landy:2008kl,Liu:2011fk,Liu:2010vn,Avitzour:2009tg,Diem:2009hc,Chen:2010bs}; however, these structures have not yet converted the absorbed light into anything useful, e.g., into an electric signal. Kirchhoff's law connects the resonant absorption with useful resonant thermal-emission properties \cite{Liu:2011fk}. The metamaterial absorption resonances can also be used for sensing \cite{Shvets:2007fu,Verellen:2009dz,Tittl:2011kx}. Furthermore, improved energy conversion in metamaterial-enhanced semiconductor-based solar cells has recently been demonstrated. These structures can also be employed as photodetectors with adjustable built-in spectral filters \cite{Knight:2011uq}, but the accessible wavelength range is fundamentally limited by the involved Schottky barrier at the semiconductor-metal interface. In this Letter, we demonstrate metal-only metamaterial bolometers that are conceptually tunable over a very large spectral range. 

\begin{figure}
\includegraphics{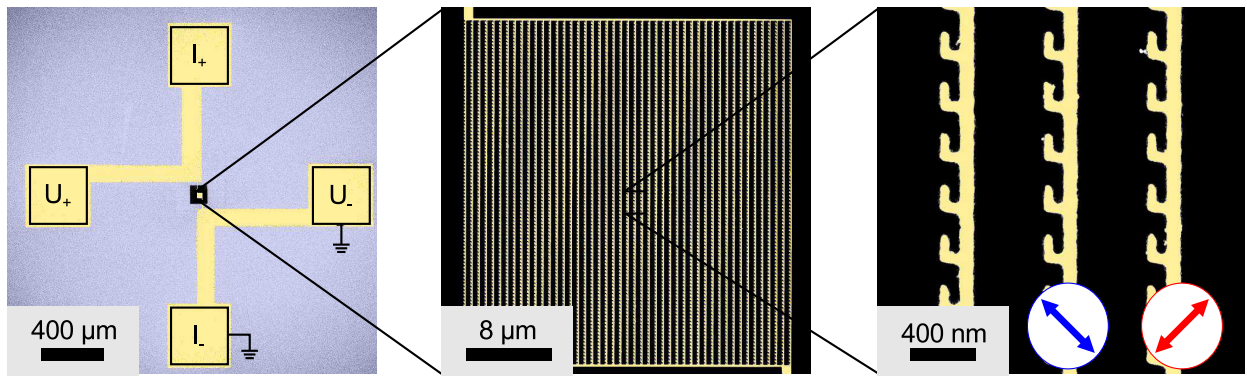}
\caption{(Color online) Metamaterial metal-based bolometer. Scanning electron micrographs of one of the fabricated devices in different magnifications as indicated by the black scale bars. The active area including the connected resonant polarization-sensitive gold absorber elements is located on a 30 nm thin SiN membrane (black). The arrows on the right-hand side indicate the orientation of the linear eigen-polarizations used in the optical experiments in Fig. 2.}
\end{figure}
To measure the temperature rise induced by resonant light absorption in metamaterials via a change in metal resistance, we use electrically connected resonant metal elements. Our corresponding design shown in Fig. 1 is based on 40 nm thin gold nanostructures on a 30 nm thin SiN membrane (with 100 \textmu m $\times$ 100 \textmu m footprint). This membrane merely serves for mechanical support, while maintaining thermal isolation. This sensor part is electrically connected to four large bond pads in the outer region, which is supported by the underlying silicon substrate. A current is injected through two of the four ports and a voltage proportional to the metamaterial resistance is detected across the other two pads. To avoid electro-migration effects \cite{Hadeed:2007qa}, we use an alternating current in the 1 kHz frequency range. This also allows for using Lock-In detection of the voltage. The entire structure has been fabricated using standard electron-beam lithography with a two-resist system to enhance undercutting, and a lift-off process. All of the below bolometer experiments have been performed at room temperature and under pre-vacuum conditions (pressure about 1 mbar) to prevent heat transport via air.

\begin{figure}
\includegraphics{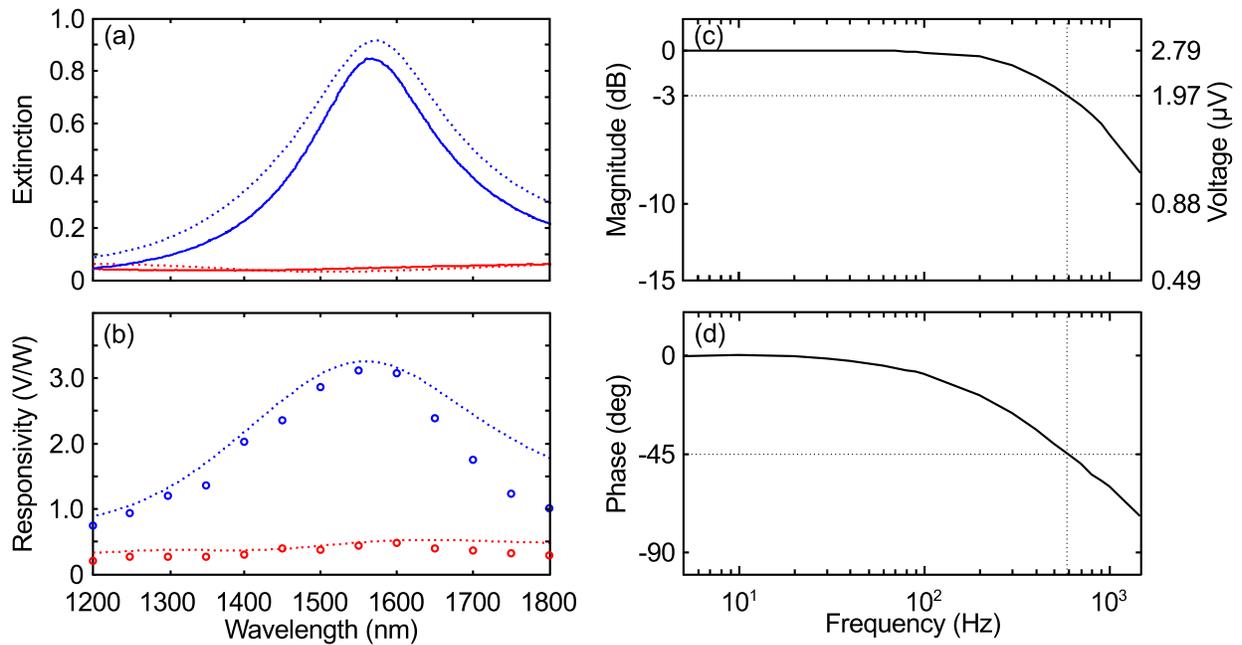}
\caption{(Color online) Optical and electrical characterization of the metamaterial bolometer.
(a), Measured (solid curves) and calculated (dashed curves) extinction spectra for the two linear polarizations indicated by the blue and red arrows in Fig. 1. (b), Measured (dots) and calculated (dashed) bolometer responsivity for an rms current of 50 \textmu A. The calculated responsivity is obtained from the calculated absorbance spectra and the measured thermal conductance $G$. (c), The thermal conductance and the bolometer time constant are obtained from the depicted measured voltage at the third harmonic of the modulation frequency measured versus modulation frequency. The bolometer time constant of 134  \textmu $s=1/(2\cdot 2\pi\cdot 596\,\rm Hz)$ results from the $-3\,\rm dB$ decay (see left vertical logarithmic scale and dashed lines). (d), Corresponding measured phase.}
\end{figure}

Electrical as well as optical characterization results are shown in Fig. 2. The thermal time constant of the bolometer is obtained from a standard \cite{Ou-Yang:1998kl} purely electrical measurement, in which the third harmonic of the bolometer voltage is recorded versus modulation frequency. The result shown in Fig. 2c,d reveals a cut-off frequency of $f=596\, \rm Hz$ which is equivalent to a time constant of about 134 \textmu s. In addition, we derive an effective thermal conductance of $G=4.85\cdot 10^{-6}\,\rm W/K$ using a temperature coefficient of resistance of $\alpha=0.0024\,\rm K^{-1}$ that we have measured independently on the same bolometer. Fig. 2b exhibits the measured responsivity versus wavelength at an rms injection current of $I=50$ \textmu A for the two orthogonal incident linear polarizations of light oriented along the two diagonals (see blue and red arrows in Fig. 1). Here, the tunable emission from an optical parametric oscillator is used as a light source that is focused onto the bolometer active area. The laser power of 50 \textmu W was kept constant for all wavelengths and polarization directions. For one linear polarization, a pronounced resonance is observed, whereas very little signal is found for the orthogonal linear polarization. This overall behavior is consistent with the independently measured extinction (negative decadic logarithm of the intensity transmittance) spectra of the device. The spectra are obtained for light impinging from the SiN membrane side normal to the membrane using a standard Fourier-transform spectrometer connected to a microscope. Our measurements have also revealed that the two diagonal linear polarizations correspond to the two eigen-polarizations of this particular structure. These overall results unambiguously show that the bolometer signal is governed by light absorption in the metal nanostructure. The corresponding bolometer voltage change is actually brought about by the resistance change of the metal. 
Unlike previous designs \cite{Maier:2009zr}, no semiconductor material is necessary in our approach. The underlying SiN membrane merely serves for mechanical support. 
\begin{figure}
\includegraphics{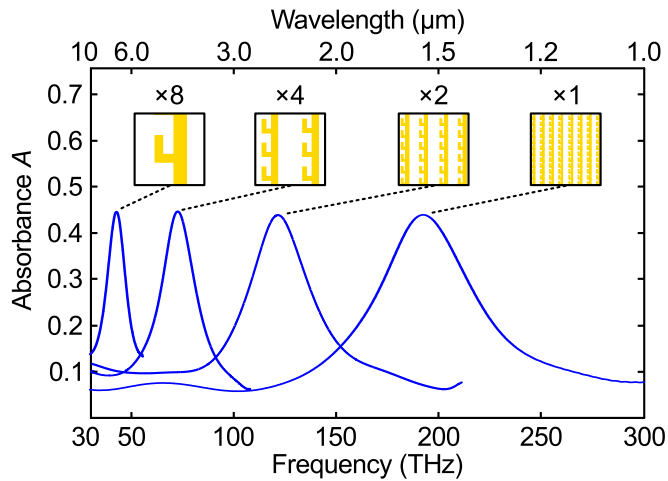}
\caption{(Color online) Calculations on bolometer tunability. For reference, the highest frequency resonance corresponds to the parameters of Fig. 2. For the other three structures, all lateral dimensions are increased in steps of factors of two (see insets), while fixing the metal thickness, hence fixing the bolometer thermal mass. Importantly, the fundamental absorber resonances can be shifted towards significantly longer wavelengths, while maintaining the peak absorbance.}
\end{figure}
The bolometer sensitivity is not only determined by the responsivity but also by the noise of the bolometer. In our laboratory setup, which uses macroscopic external wiring and external circuitry, the electric noise is entirely governed by extrinsic effects. Fundamentally, upon integrating suitable electronics on chip, the noise of metal bolometers is known to be determined only by the Brownian motion of metal electrons, i.e., by so-called Johnson noise. For the resistance of our bolometer of $R=300\,\Omega$ and room-temperature vacuum operation, Johnson noise translates into a noise equivalent power of $6.7\cdot 10^{-10}\, \rm W/\sqrt{Hz}$. 
\begin{figure}
\includegraphics{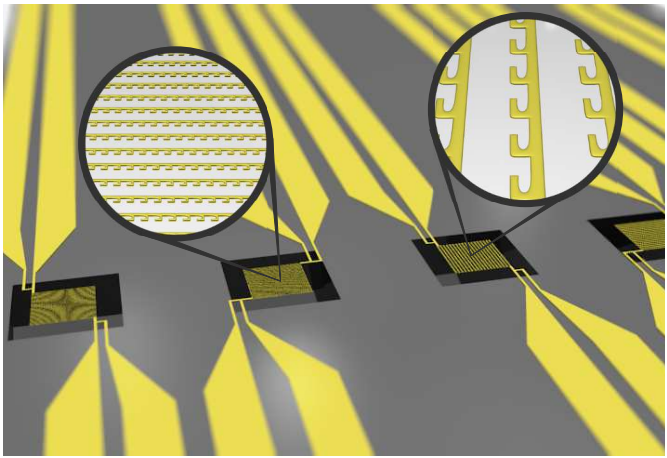}
\caption{(Color online) Envisioned bolometer array architecture allowing for integrated broadband spectroscopy and polarimetry without the need for any external dispersive element or any polarizer.}
\end{figure}

To support our above interpretation of the experimental results and to evaluate the spectral tuning capabilities of our bolometer approach, we have performed additional numerical calculations using the commercial software package CST Microwave Studio, which is based on a finite-integration time-domain method. The gold is described by the free-electron Drude model with plasma frequency $\omega_{\rm pl}=1.37\cdot 10^{16}$ rad/s and collision frequency $\omega_{\rm col}=2.2\cdot 10^{14}$ rad/s. The geometrical parameters are taken from the electron micrograph shown in Fig. 1, the gold thickness is 40 nm as in the experiment. The refractive index of the 30 nm thin SiN membrane is taken as $n=1.96$. Fig. 2a,b show the calculated extinction and responsivity spectra together with the experimental results. The calculated responsivity spectra are obtained from the numerically calculated absorbance spectra, $A$, and the expression for the responsivity $\Re\approx A  I R\alpha/G$ \cite{LIDDIARD:1984dz}.
Both the spectrally resonant behavior and the polarization dependence are qualitatively well reproduced. Minor quantitative differences are likely due to the simplifications and/or to experimental imperfections.\\
This good qualitative agreement can be taken as the starting point for evaluating the tunability of the resonant bolometer response, which is required for implementing broadband spectrometers. In principle, owing to the scalability of the Maxwell equations, the metamaterial resonances can simply be shifted towards longer wavelengths by proportionally increasing the size of all features. Here we only scale the lateral features and keep the metal thickness constant in order to fix the bolometer thermal mass. In Fig. 3, we show corresponding calculations. The lateral features are increased in steps of factors of two. Clearly, comparable peak absorbance values are achieved for light polarized along the respective resonant eigen-polarization (see blue arrow in Fig. 1). Due to the lower metal damping at longer wavelengths, however, the resonances also become significantly narrower. This aspect is beneficial when aiming at realizing integrated bolometer spectrometers as the resonance linewidth obviously limits the achievable spectral resolution. Overall, a tuning between roughly  1 \textmu m and 10 \textmu m wavelength is easily possible along these lines.

In conclusion, we have demonstrated individual metamaterial metal-based bolometer elements. On this basis, we envision integrated broadband metamaterial metal-based bolometer spectrometer arrays (Fig. 4). Adding rotated elements would simultaneously allow for analyzing the polarization state of light without the need for any external dispersive element or any external polarizer.

We acknowledge support by the DFG-Center for Functional Nanostructures (CFN) via subproject A1.5. The project METAMAT is supported by the Bundesministerium f\"ur Bildung und Forschung (BMBF). The PhD education of F.B.P.N. and J.K.G is embedded in the Karlsruhe School of Optics \& Photonics (KSOP).

\providecommand*\mcitethebibliography{\thebibliography}
\csname @ifundefined\endcsname{endmcitethebibliography}
  {\let\endmcitethebibliography\endthebibliography}{}

\end{document}